\documentclass[12pt]{article}
\usepackage{graphicx}
\usepackage{hepnames}
\usepackage{amsmath}
\usepackage{adjustbox}
\usepackage{enumitem}
\usepackage{hyperref}
\usepackage{caption}
\usepackage{subcaption}
\usepackage{comment}

\newcommand\pubnumber{DPF2015-342}
\newcommand\pubdate{\today}

\def\yale{Department of Physics\\
Wright Laboratory\\Yale University\\ New Haven, CT 06511}
\def\support{\footnote{Presenter: \href{mailto:danielle.norcini@yale.edu}{danielle.norcini@yale.edu}}}

\def\Title#1{\begin{center} {\Large #1 } \end{center}}
\def\Author#1{\begin{center}{ \sc #1} \end{center}}
\def\Address#1{\begin{center}{ \it #1} \end{center}}

\newcommand\pubblock{\rightline{\begin{tabular}{l} \pubnumber\\
         \pubdate  \end{tabular}}}
\newenvironment{Abstract}{\begin{quotation}  }{\end{quotation}}
\newenvironment{Presented}{\begin{quotation} \begin{center} 
             PRESENTED AT\end{center}\bigskip 
      \begin{center}\begin{large}}{\end{large}\end{center} \end{quotation}}
\def\Acknowledgments{\bigskip  \bigskip \begin{center} \begin{large}
             \bf ACKNOWLEDGMENTS \end{large}\end{center}}

\def\beq{\begin{equation}}
\def\eeq#1{\label{#1}\end{equation}}
\def\eeqn{\end{equation}}

\def\beqa{\begin{eqnarray}}
\def\eeqa#1{\label{#1}\end{eqnarray}}
\def\eeqan{\end{eqnarray}}

\let\bar=\overbar

\def\Dslash{\not{\hbox{\kern-4pt $D$}}}
\def\dslash{\not{\hbox{\kern-2pt $\del$}}}

\def\msb{{\bar{\ssstyle M \kern -1pt S}}}

\begin{document}
\begin{titlepage}
\pubblock

\vfill
\Title{Development of PROSPECT detectors for precision antineutrino studies}
\vfill
\Author{{Danielle Norcini}\support}
\Address{\yale}
\Author{on behalf of the PROSPECT collaboration}
\vfill
\begin{Abstract}
\noindent PROSPECT, the Precision Reactor Oscillation and Spectrum Experiment, will use two segmented detectors positioned 7-20~m from the High Flux Isotope Reactor (HFIR) at Oak Ridge National Laboratory to measure the $^{235}$U antineutrino spectrum and perform a search for short-baseline oscillations as a signature of eV-scale sterile neutrinos. PROSPECT has developed $^{6}$Li-loaded liquid scintillator detectors for efficient identification of reactor antineutrinos and has measured reactor and cosmogenic backgrounds in the HFIR reactor building. Multiple test detectors have been built, operated, and characterized at HFIR and elsewhere to understand the optical performance of the scintillator and pulse-shape discrimination capabilities for enhanced background rejection. The results from this R\&D effort are discussed, in the context of the design and physics potential of PROSPECT.
\end{Abstract}
\vfill
\begin{Presented}
DPF 2015\\
The Meeting of the American Physical Society\\
Division of Particles and Fields\\
Ann Arbor, Michigan, August 4--8, 2015\\
\end{Presented}
\vfill
\end{titlepage}
\def\thefootnote{\fnsymbol{footnote}}
\setcounter{footnote}{0}

\section{Introduction}
\label{sec:intro}
The study of \Pagne emitted by nuclear reactors has played a vital role in the field of neutrino physics. Oscillations among the 3 massive neutrino flavors were established by atmospheric, solar and reactor experiments. Current results from accelerator experiments have also verified the 3-flavor mixing scheme. However, recent reactor experiments have produced results that disagree with predictions, namely, the observation of a $\sim$5\% deficit in the absolute flux \cite{{Mueller:2011nm},{Huber:2011wv}} and deviations in the \Pagne energy spectrum within the 4-7~MeV region \cite{{Seon-HeeSeofortheRENO:2014jza},{Abe:2014bwa},{An:2015nua}}. The flux deficit, referred to as the``reactor antineutrino anomaly'', may be explained by the addition of light sterile neutrinos \cite{{abazajian2012light},{Kopp2013vaa}}, which would indicate physics beyond the Standard Model. The spectral deviations measured by the $\theta_{13}$ experiments may point to an incomplete understanding in reactor fission models or other new physics \cite{{Hayes:2015yka}, {Dwyer:2014eka}, {Sonzogni:2015aoa}}.

PROSPECT, the Precision Reactor Oscillation and Spectrum Experiment \cite{Ashenfelter:2013oaa}, is a short-baseline reactor neutrino experiment that will perform an oscillation search for possible sterile neutrinos with mass splittings $\mathcal{O}(1~\mathrm{eV^{2}})$ and make a precision measurement of the $^{235}$U reactor \Pagne spectrum from the High Flux Isotope Reactor (HFIR) at Oak Ridge National Laboratory (ORNL). PROSPECT is a phased experiment and will begin data-taking with the Phase I antineutrino detector (AD). The AD will enclose a total active mass of 3~tons and have variable baseline coverage of 7-12~m from the compact reactor core. The Phase II AD will be constructed for additional sensitivity and statistics, with mass $\mathcal{O}(10~\mathrm{ton})$ and cover baselines of 16-20~m. 

Both the Phase I and II ADs are $^{6}$Li-loaded liquid scintillator (LiLS) detectors that are optically segmented into rectangular cells with double-ended photomultiplier tube (PMT) readout. Neutrinos are detected through the inverse beta decay (IBD) interaction: $\mathrm{\Pagne + p \rightarrow e^{+} + n.}$ Antineutrinos from $\mathrm{\beta}$ decays of $^{235}$U fission products interact with protons in the LiLS, consequently emitting a e$^{+}$ and a neutron. Due to the large mass difference between the product particles, the e$^{+}$ carries most of the neutrino kinetic energy. The kinetic energy is deposited in the LiLS, along with the energy from two annihilation $\gamma$-rays. Such an event is labeled the ``prompt'' signal, as it occurs almost instantaneously, with characteristic energy of 1-10~MeV. This prompt signal holds the information required to make the precision spectral measurement. The neutron, with small kinetic energy, scatters around the LiLS for $\sim$40$~\mathrm{\mu}$s before capturing on a $^{6}$Li nucleus: $\mathrm{n + ^{6}Li \rightarrow \alpha + t + 4.78~MeV.}$ Subsequent decay of the nucleus creates an $\alpha$-triton pair. Since the $\mathrm{\alpha}$ and triton carry electric charge, these heavy particles ionize the LiLS and this ``delayed'' signal is registered at the quenched neutron capture energy of $\sim$0.6~MeV$_{ee}$, or electron equivalent MeV. Based on the coincidence of the prompt and delayed signals, stringent cuts can be made on energy, time, and geometry to identify IBD events within PROSPECT. 

\section{PROSPECT Phase I detector}
Over the last year, the PROSPECT collaboration has conducted an extensive R\&D program focused on designing the Phase I detector and understanding its performance potential. The AD, with a moveable baseline ranging 7-12~m from the HFIR core, will search the reactor anomaly parameter space for possible sterile neutrinos at 3$\sigma$ in one calendar year with a reactor up-time of 41\%. It will also make a precision measurement of the $^{235}$U spectrum with energy resolution of 4.5\%/$\mathrm{\sqrt{E}}$. 

The detector consists of 3~tons LiLS, which will be divided into a 10x12 segmented array via low-mass optical separators to allow for interaction topology cuts and observations of oscillations between segments. Each segment will have double-ended PMT readout and share hollow support rods with its nearest neighbors for calibration access. A multi-layer passive shield surrounding the exterior of the detector will be utilized to suppress cosmogenic and reactor-related backgrounds.

\begin{figure}[htb]
\centering
\includegraphics[width=.9\textwidth]{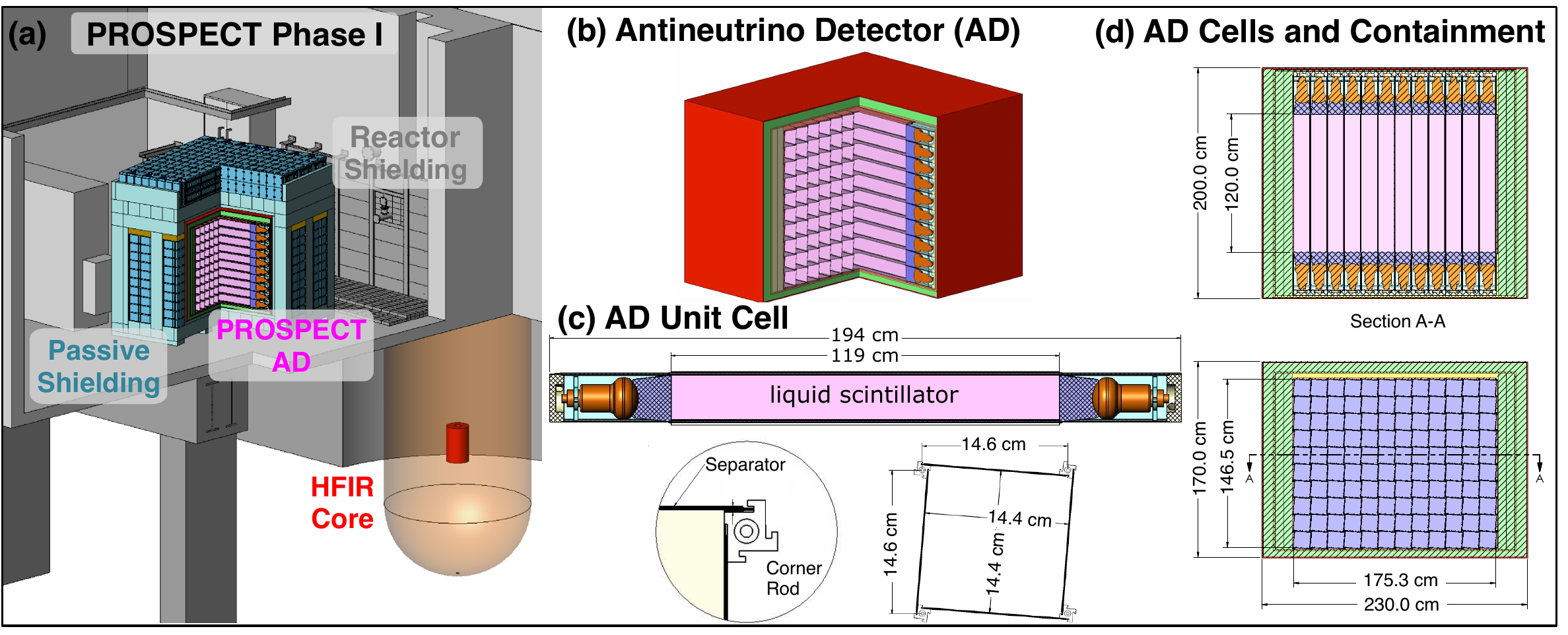} 
\caption{(a) The Phase I AD in the near position at HFIR. (b) A cutaway diagram of the AD revealing unit cell array structure. (c) The unit cell structure. (d) Inner and outer dimensions of the AD.} 
\label{fig:expRealization}
\end{figure}

\noindent To understand the detector components, PROSPECT has developed several prototypes with a specified purpose, as listed in Table \ref{tab:prototype}. With the creation of each prototype, background Monte Carlo and detector performance models were produced and validated against the data.

\begin{table}[th!]
\begin{center}
\begin{adjustbox}{width=\textwidth}
\begin{tabular}{|c|c|c|c|l|}  
\hline
 Prototype & Volume & Scintillator & Location & Purpose \\ \hline
 PROSPECT-0.1  &  100~mL   &  EJ-309  &  HFIR & characterize liquid scintillator properties  \\
 PROSPECT-0.1  &  100~mL   &  $^{6}$Li EJ-309  &  Yale & characterize liquid scintillator properties  \\
 PROSPECT-2  &  1.8~L   &  $^{6}$Li EJ-309  &  HFIR & understand near-surface backgrounds \\
 PROSPECT-20  &  23~L   &  EJ-309  &  Yale & characterize optical transport properties of a segment \\
 PROSPECT-20  &  23~L   &  $^{6}$Li EJ-309  &  HFIR, Yale & operate full-size segment in reactor environment\\
\hline
\end{tabular}
\end{adjustbox}
\caption{PROSPECT prototype detectors.}
\label{tab:prototype}
\end{center}
\end{table}

\section{Background mitigation techniques}
With minimal overburden, PROSPECT will face both cosmogenic and reactor-related backgrounds at HFIR in the Phase I near position. The major sources of background include correlated cosmogenic fast neutrons, correlated multiple neutron captures on $^{6}$Li from cosmic showers, and uncorrelated reactor $\mathrm{\gamma}$-rays. 

As detailed in \cite{Ashenfelter:2015tpm}, PROSPECT has conducted a detailed evaluation of the three research HEU reactor facilities in the US: the Advanced Test Reactor (ATR) at Idaho National Laboratory (INL), HFIR at ORNL, and the National Bureau of Standards Reactor (NBSR) at the National Institute of Standards and Technology (NIST). The sources and distributions of the reactor and cosmogenic backgrounds were characterized at each site. All sites were determined viable for the PROSPECT experiment. Based on the detailed background studies, HFIR was chosen due to logistics considerations, access, and operations. Further background measurements at HFIR suggested a set of background reduction techniques devised to allow for a signal-to-background ratio $>$1 at HFIR:

\begin{itemize}[topsep=0pt,itemsep=-1ex,partopsep=1ex,parsep=1ex]
	\item employ a multi-layer passive shield surrounding the detector to suppress neutrons and $\mathrm{\gamma}$-rays
	\item use time coincidence of PMT hits that correspond to the IBD signature
	\item rely on particle identification from pulse-shape discrimination (PSD)
	\item utilize detector segmentation for topology cuts and efficient fiduciailization  
\end{itemize}

Detector design considerations along with analysis strategies are used to control and reduce the backgrounds in PROSPECT. As discussed in Section \ref{sec:intro}, the Phase I detector will be optically segmented. This enables the implementation of topological cuts in analysis, due to the localized nature of IBD events in LiLS. The segments also allow for simple fiducialization, permitting the outer volume to act as $\mathrm{\gamma}$-ray catcher. The exterior of the detector has been designed to accommodate passive shielding with borated-polyethylene to suppress fast neutrons and includes a targeted lead wall facing the reactor to reduce reactor $\gamma$-rays.

\begin{figure}[t!]
\centering
	\begin{subfigure}[b]{0.5\textwidth}
	\centering
	\includegraphics[width=.9\textwidth]{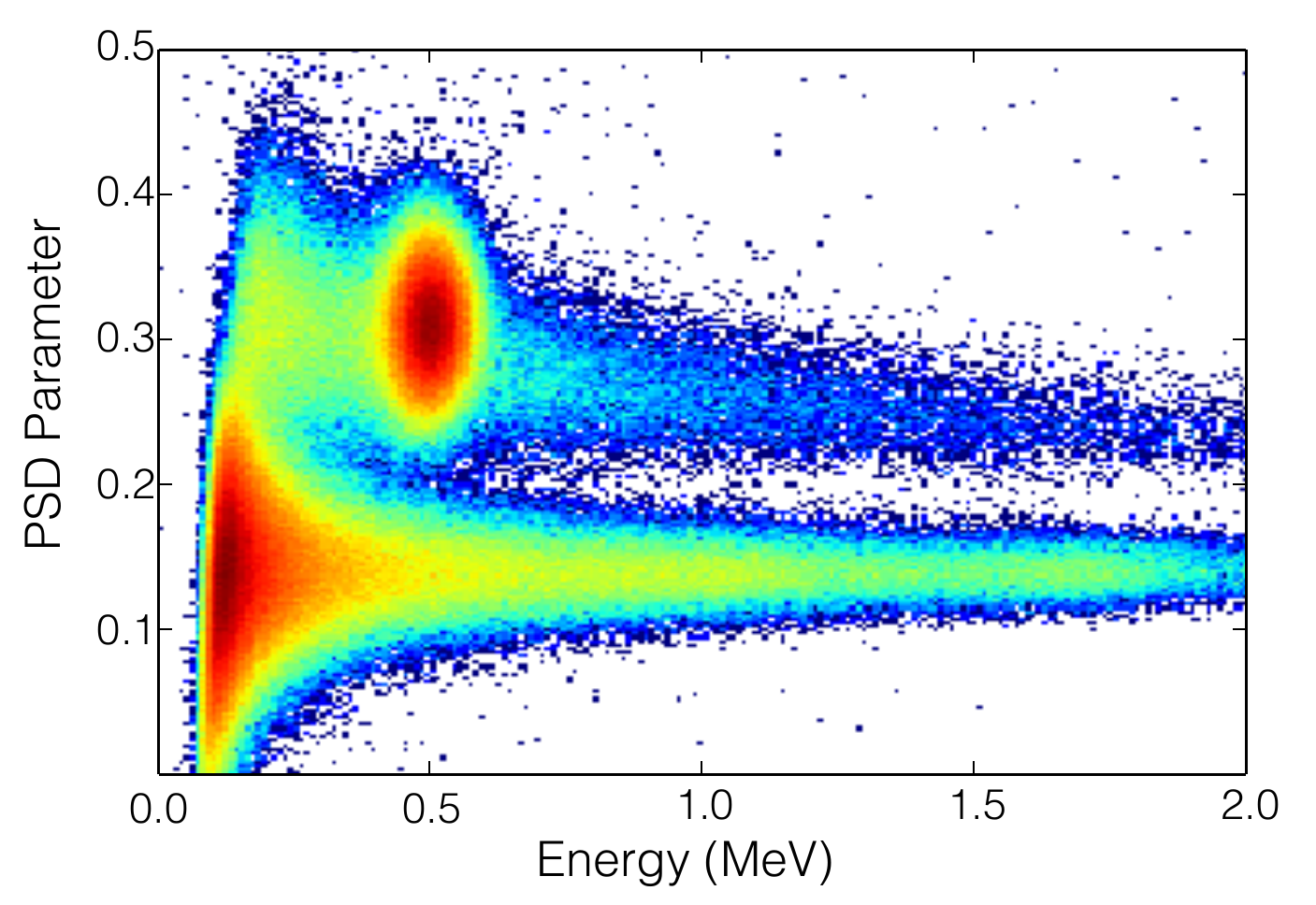}
	\subcaption{ }
	\label{fig:psd_energy}
	\end{subfigure}%
	\begin{subfigure}[b]{0.5\textwidth}
	\centering
	\includegraphics[width=.74\textwidth]{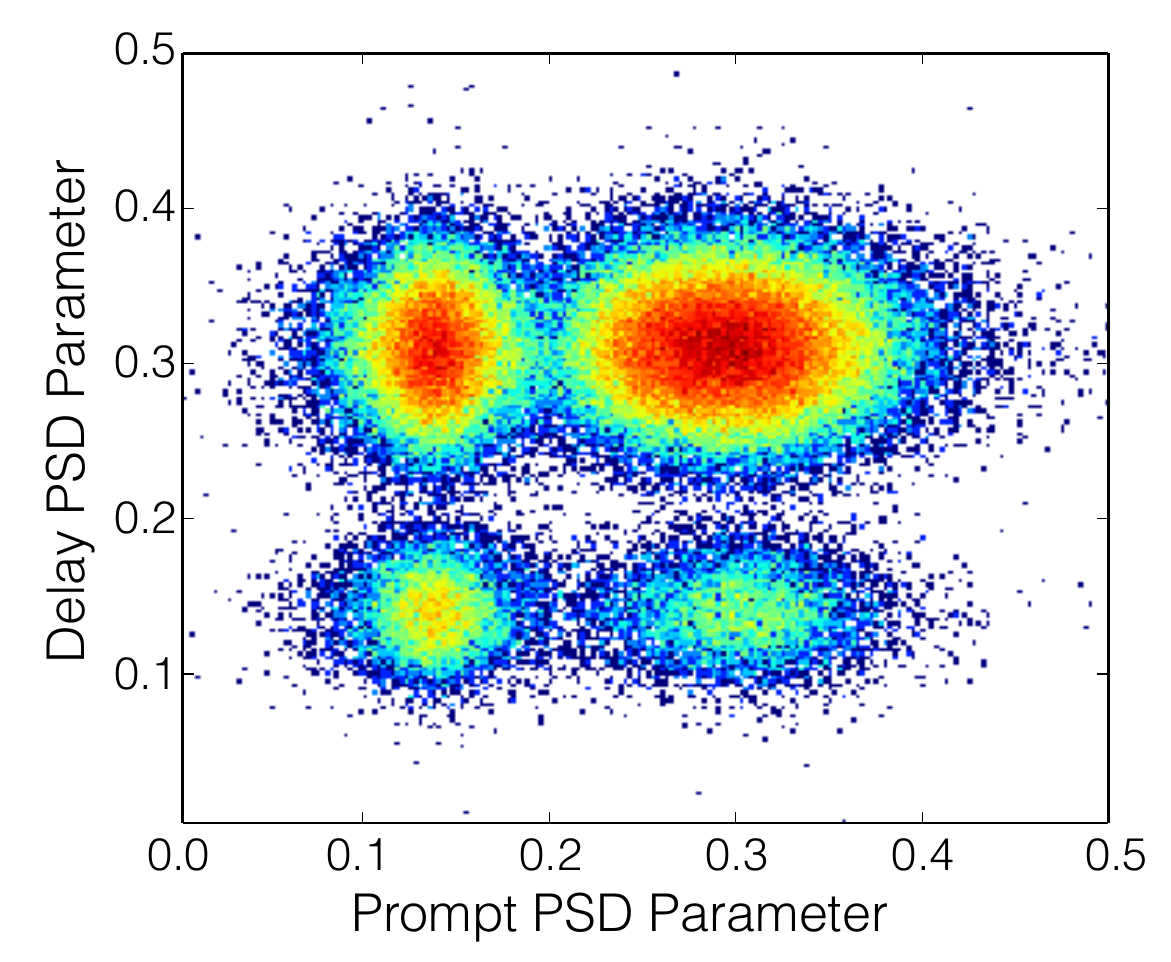}
	\subcaption{ }
	\label{fig:psd_cuts}
	\end{subfigure}
\caption{(a) PSD parameters and energies of measured $^{252}$Cf spontaneous fission $\gamma$-rays and neutrons in PROSPECT-2. The neutron capture peak is prominent at low energies (0.5-0.7~MeV) and high PSD. There is a large separation between gamma-like particles (bottom band) and neutron-like particles (top band).  (b) PSD parameter values for prompt and delayed signals in detected time-coincident triggers in PROSPECT-2.  Separation between IBD-like coincidences (top left), accidental gamma coincidences (bottom left), and fast- or multiple-neutron related coincidences (top right) is clearly visible \cite{Ashenfelter:2013oaa}.}
\end{figure}

In analysis, backgrounds from fast neutrons and $\gamma$-rays can be eliminated through pulse-shape discrimination (PSD) cuts on the prompt and delayed events. Particles resulting from electromagnetic interactions can be identified as ``gamma-like'', whereas hadronic particles are labelled ``neutron-like''. As seen in Figure \ref{fig:psd_energy}, two bands are clearly visible, one indicating the presence of gamma-like particles, and the other neutron-like. The neutron capture peak on $^{6}$Li is also prominent. The PSD parameter used in this analysis can be expressed as

\begin{equation}
PSD = \frac{Q_{tail}}{Q_{full}},
\end{equation}

\noindent where Q$_{tail}$ is the amount of charge found in the PMT pulse tail and Q$_{full}$ is the total charge. With energy cuts corresponding to prompt and delayed signals, two-dimensional plots of the prompt PSD parameter and the delayed PSD parameter can be created. It is then possible to identify fast neutron backgrounds with a neutron-like prompt and delay signal. Similarly for accidental $\mathrm{\gamma}$-rays, there is a gamma-like prompt and delay PSD signal, as illustrated in Figure \ref{fig:psd_cuts}. Rejecting these candidates results in the remaining IBD-like candidates, with gamma-like prompt and neutron-like delay signals.

The ability to observe clean PSD signatures for background reduction heavily depends on the scintillator properties. PROSPECT has developed a lithium-loading technique that is compatible with EJ-309, a common commercial liquid scintillator with excellent PSD. The performance of this novel scintillator cocktail has been fully characterized with the PROSPECT-0.1 prototype. Results indicate a light yield of 8200 photons/MeV and PSD figure-of-merit at (n,Li) of 1.79, which is defined as
\begin{equation}
FOM = \frac{\lvert\mu_{1}-\mu_{2}\rvert}{FWHM_{1}+FWHM_{2}}
\end{equation}
\noindent where $\mathrm{\mu_{1}}$, $\mathrm{\mu_{2}}$ represent the mean of the gamma-like and neutron-like distributions, and FWHM$_{1}$, FWHM$_{2}$ the full-width-half-max of each distribution. The LiLS has demonstrated excellent light yield for the required energy resolution, optimized PSD for particle identification, and a distinguishable neutron capture peak.

\section{PROSPECT-20: Segment response studies}
PROSPECT-20 is an acrylic (15~cm x 15~cm x 1~m) test cell designed to study of the optical behavior of a full-sized AD segment. Light collection capabilities, PSD performance, and PMT readout type were explored in detail \cite{Ashenfelter:2015aaa}.

\begin{figure*}[htb!pb]
\centering
\includegraphics[trim={0cm 0cm 0cm 0cm},clip,width=0.65\textwidth]{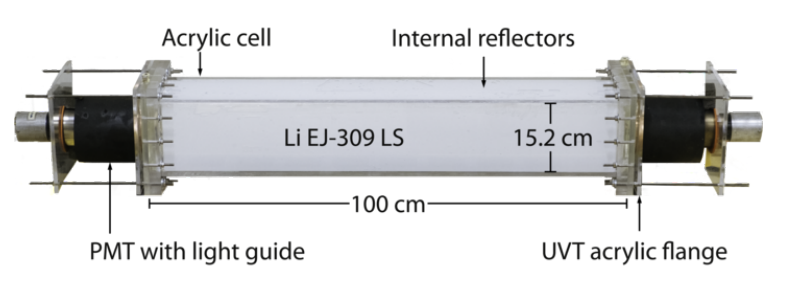}
\caption{The meter-long \textsc{PROSPECT-20} prototype.}
\label{fig:p20_cell}
\end{figure*}

The detector, shown in Figure \ref{fig:p20_cell}, is equipped with Hamamatsu R6594 PMTs and internal reflectors based on the low-mass optical separators design. When filled with undoped EJ-309, a light collection of 841$\pm$17~photoelectrons~(PE)/MeV was measured with excellent PSD performance. Quantifying the rejection capabilities around the expected region (0.5-0.7~MeV) for neutron capture ($\sim$0.6~MeV$_{ee}$ in LiLS), the detector demonstrated the ability to reject 99.99\% $\mathrm{\gamma}$-rays, while maintaining 99.9\% of the neutron events. It was also shown that double-ended PMT readout  is essential for uniform optical collection, enhancement of PSD, and effective axial position reconstruction. 

\begin{figure}[t!]
\centering
	\begin{subfigure}[b]{0.5\textwidth}
	\centering
	\includegraphics[width=0.7\textwidth]{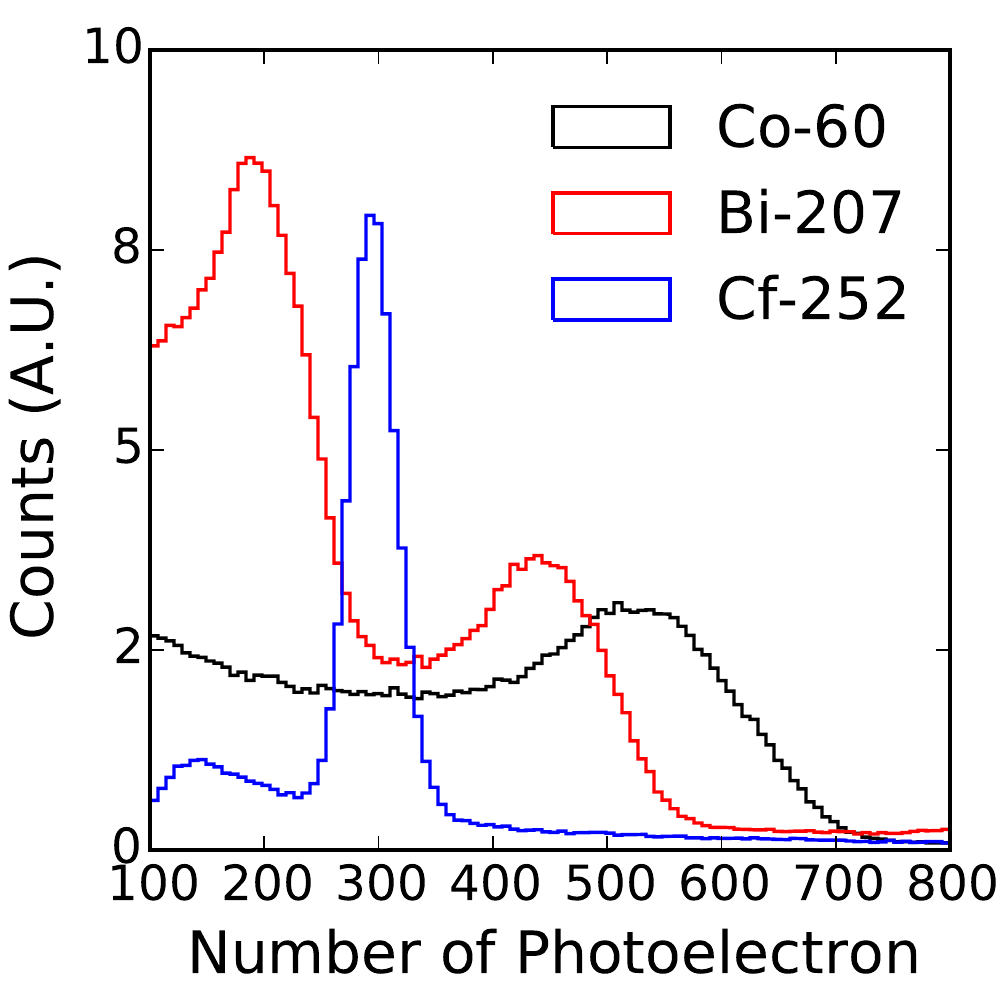}
	\subcaption{ }
	\label{fig:p20_spectrum}
	\end{subfigure}%
	\begin{subfigure}[b]{0.5\textwidth}
	\centering
	\includegraphics[width=.7\textwidth]{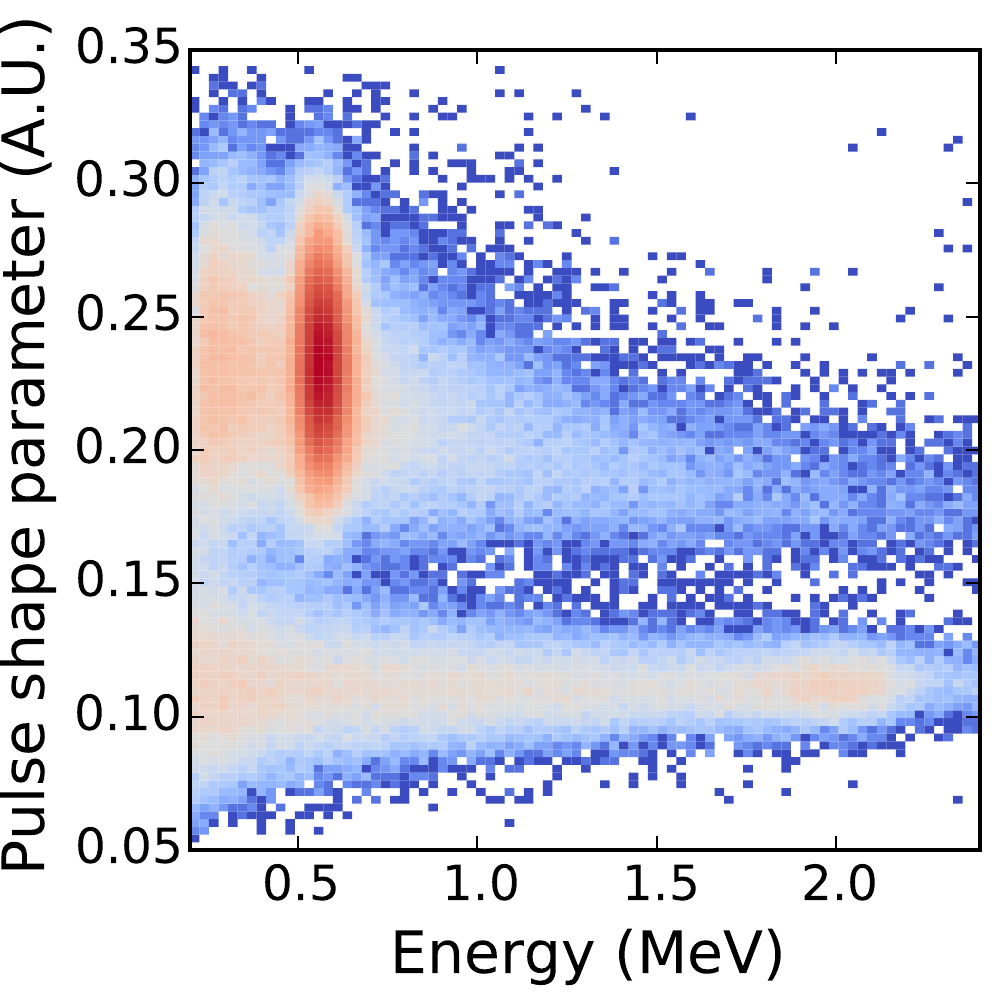}
	\subcaption{ }
	\label{fig:p20_psd}
	\end{subfigure}
\caption{(a) Measured photoelectron spectra from three sources. The Compton edge of $^{60}$Co and $^{217}$Bi $\mathrm{\gamma}$-rays and the quenched neutron capture peak from $^{252}$Cf neutrons indicate a collection efficiency of $522\pm16$~PE/MeV. (b) PSD versus energy for $^{252}$Cf.}
\end{figure}

To more accurately understand the performance of the LiLS in a realistic geometry, PROSPECT-20 was filled with $^{6}$Li EJ-309. Using $^{60}$Co, $^{207}$Bi, and $^{252}$Cf as sources with radiations in the range 0.38-2.0~MeV, an average light collection of 522$\pm$16~PE/MeV was achieved, as shown in Figure \ref{fig:p20_spectrum}. To reach the target PROSPECT AD resolution of 4.5\%/$\mathrm{\sqrt{E}}$, a light collection of 500~PE/MeV is required, which is achieved in PROSPECT-20. The PSD performance of the LiLS remains excellent (Figure \ref{fig:p20_psd}), exhibiting a prominent neutron capture peak preserving 99.9\% of the neutron events and rejecting a similar fraction of the $\mathrm{\gamma}$-ray signal. The results from PROSPECT-20 indicate that the PROSPECT AD segments will meet the performance criteria needed for a short-baseline sterile neutrino search and precision spectral measurement.


\section{Conclusions}
PROSPECT will perform an oscillation-based search for light sterile neutrinos and make a precise measurement of the $^{235}$U spectrum at short-baselines from HFIR. The R\&D program put forth by the collaboration has resulted in multiple prototype detectors to validate detector performance and simulations models, developed a detailed understanding of the near-surface backgrounds at HFIR with accompanying background mitigation techniques, and demonstrated the light collection and PSD performance required for the Phase I detector. With Phase I, PROSPECT will probe the favored region of the reactor anomaly parameter space within 1 calendar year at 3$\mathrm{\sigma}$ and $>$5$\mathrm{\sigma}$ in 3 years.

\Acknowledgments
This material is based upon work supported by the U.S. Department of Energy Office of Science. Additional support for this work is provided by Yale University and the Illinois Institute of Technology. We gratefully acknowledge the support and hospitality of the High Flux Isotope Reactor and the Physics Division at Oak Ridge National Laboratory, managed by UT-Battelle for the U.S. Department of Energy.

\end{document}